\documentclass[12pt]{article}
\usepackage[margin=1.1in]{geometry}
\usepackage{amsmath,amssymb,amsfonts}
\usepackage{graphicx}
\usepackage{booktabs}
\usepackage[colorlinks=true,linkcolor=blue,citecolor=blue,urlcolor=blue]{hyperref}
\usepackage{authblk}
\usepackage{float}
\usepackage[title]{appendix}

\newcommand{\Mpl}{M_{\rm pl}}
\newcommand{\mKK}{m_{\rm KK}}
\newcommand{\mDM}{m_{\rm DM}}

\begin{document}

\begin{titlepage}
\begin{center}

\vspace*{1.4cm}

{\LARGE \bf The Phenomenology of the Dark Dimension\\[2mm] and Non-Supersymmetric Strings}

\vspace{1.4cm}

{\large Cumrun Vafa}

\vspace{0.5cm}

{\it Jefferson Physical Laboratory, Harvard University,\\ Cambridge, MA 02138, USA}

\vspace{1.6cm}

\begin{abstract}
\noindent
Swampland principles, combined with the smallness of dark energy $\Lambda$, lead to the dark dimension scenario, with one extra dimension of micron size. We review this scenario and its predictions: KK graviton dark matter (or the KK tower of
the quintessence field into which the gravitons convert), neutrino masses tied to $\Lambda^{1/4}$ via a bulk $B-L$ gauge symmetry, QCD axion decay constant $f_a\sim \Lambda^{1/12}\sim 10^9-10^{10}$ GeV at the  5d Planck scale $M_5$, connecting electroweak hierarchy to $\Lambda^{1/6}$, and a relation between the GUT and weak scales with $M_5$ as the geometric mean. The dark dimension also provides a concrete realization of an evolving dark sector: the evolution of its diameter correlates the variation of dark energy with that of the dark matter mass, fitting the recent DESI DR2 (+CMB+supernovae) data while avoiding the phantom crossing that the $w_0w_a$CDM parametrization would suggest---the apparent phantom behavior instead arises from the evolving dark matter mass. The fits suggest that the dark energy field sits in the interior of moduli space, near an unstable critical point. Motivated by this, we review recent progress on non-supersymmetric string vacua: isolated critical points protected by duality symmetries, in non-geometric flux compactifications and in Scherk--Schwarz compactifications of M-theory, and an emerging duality web for non-supersymmetric strings. This note is based on two talks given by the author at String Phenomenology 2026 (Paris) and Strings 2026 (Shanghai).
\end{abstract}

\end{center}
\end{titlepage}

\tableofcontents
\newpage

\section{Introduction}
\label{sec:intro}

Some of the deepest puzzles in our universe concern the dark sector and the structure of scales in particle physics: What is the nature of dark energy and dark matter? Why are neutrino masses so small, and why do they happen to coincide with the dark energy scale? Is there a QCD axion, and what fixes its decay constant? Do the gauge couplings unify, and if so how is the unification scale related to the weak scale? To this list one may add more recent puzzles suggested by cosmological observations, in particular the hints from DESI and DES that dark energy may not be a cosmological constant but is evolving.

The Swampland program (see \cite{Vafa:2005ui,Arkani-Hamed:2006emk,Ooguri:2006in} for the early works and \cite{Palti:2019pca,vanBeest:2021lhn,Grana:2021zvf,Agmon:2022thq} for more recent reviews) aims to delineate which low-energy effective field theories can be consistently coupled to quantum gravity. Remarkably, when combined with a small number of experimental facts, most importantly the observed smallness of the dark energy,
\begin{equation}
\Lambda \sim 10^{-122}\,\Mpl^4\,,
\label{eq:cc}
\end{equation}
Swampland principles lead to an essentially unique corner of the quantum gravity landscape: the \emph{dark dimension scenario}~\cite{Montero:2022prj}, with one extra dimension of micron size. The first part of this paper reviews this scenario and its predictions, based on works with my collaborators. There are also other approaches to the phenomenology of this scenario (see e.g.~\cite{anchordoqui2024landscapeswamplandextradimensions} and references therein), which we will not review here.

The predictions of the scenario tie together, in a single framework, the puzzles listed above: the Kaluza--Klein (KK) tower of gravitons in the dark dimension provides the dark matter~\cite{Gonzalo:2022jac,Law-Smith:2023czn,Obied:2023clp}; a bulk $B-L$ gauge symmetry naturally requires three right-handed neutrinos propagating in the bulk and explains $m_\nu\sim \Lambda^{1/4}$~\cite{Montero:2025rvz}; the Weak Gravity Conjecture ties the QCD axion decay constant to the five-dimensional Planck scale~\cite{Gendler:2024adn}; and winding versus momentum modes of a Planckian string on the Standard Model (SM) brane relate the GUT scale to the weak scale~\cite{Heckman:2024oot}. Moreover, the de Sitter conjecture~\cite{Obied:2018sgi} and the Trans-Planckian Censorship Conjecture (TCC)~\cite{Bedroya:2019snp} suggest that dark energy is evolving. An economical realization where the radius of the dark dimension itself is evolving, so that the dark matter mass varies along with the dark energy was anticipated \cite{Agrawal:2019dlm} and fits \cite{Bedroya:2025fwh} the recent DESI DR2 data (in combination with CMB and supernovae datasets) as well as the phenomenological $w_0w_a$CDM parametrization does, while providing a microscopic origin for the apparent phantom behavior (see also~\cite{Vafa:2024fpx} for a review). Strikingly, all the mysterious hierarchies of scales in our universe can then be organized as simple powers of the dark energy $\Lambda$, updating and suggesting a resolution to Dirac's large-number puzzle.  Moreover, as we point out in Appendix~\ref{app:phitower}, the fits with cosmological data imply
that the field driving dark energy is essentially a massless 5d field, and therefore comes
with its own KK tower in the dark dimension; this tower is a natural dark
matter candidate, fed by the conversion of the KK gravitons (which also
sidesteps the obstruction of \cite{Lee:2026kkg,Langhoff:2026wiz} to the
graviton cascade if we restrict to two derivative action) and offers an alternative realization of the evolving dark sector in which the dark dimension itself stays fixed.

The fits to the data have a further lesson: the effective logarithmic slope of the dark energy potential predicted as well as fitting  to order one in Planck units is smaller than its allowed asymptotic minimum.  This suggests that the field controlling dark energy sits in the \emph{interior} of moduli space, perhaps close to an unstable critical point. Since our universe has no low-energy supersymmetry and a tiny positive potential, this sharpens a central challenge for string theory: understanding non-supersymmetric string vacua and, in particular, critical points of the potential in the interior of field space.   This motivates focusing on the construction of novel string landscapes, instead of meta-stable dS, of unstable dS.  The second part of this paper reviews recent progress in this direction: symmetry-protected isolated critical points in non-geometric flux compactifications~\cite{Becker:2006ks,Rajaguru:2024emw,Chen:2025rkb,BentoMontero:2025}, isolated critical points of Scherk--Schwarz compactifications of M-theory identified via unbroken duality groups~\cite{Baykara:2026scherk}, and an emerging duality web for non-supersymmetric strings~\cite{Baykara:2026web}, including evidence for dualities relating the bosonic string to superstrings. Non-supersymmetric string vacua are beginning to get demystified.

The organization of this paper is as follows. In Section~\ref{sec:dd} we review the dark dimension scenario. Sections~\ref{sec:dm}--\ref{sec:gut} discuss its predictions for dark matter, neutrinos and $B-L$, the QCD axion, and grand unification. Section~\ref{sec:fds} discusses the evolving dark sector and its confrontation with recent data, and Section~\ref{sec:hierarchy} collects the hierarchies of scales as powers of $\Lambda$. Section~\ref{sec:nonsusy} motivates the study of non-supersymmetric vacua; Section~\ref{sec:symmetries} reviews symmetry-protected critical points in non-geometric compactifications; Section~\ref{sec:ss} discusses Scherk--Schwarz compactifications of M-theory and the emerging duality web for non-supersymmetric strings. We conclude in Section~\ref{sec:conclusions}. Appendix~\ref{app:phitower} discusses the KK tower of the quintessence field as a dark matter candidate.

\section{The Dark Dimension Scenario}
\label{sec:dd}

The starting point is the Distance/Duality Conjecture~\cite{Ooguri:2006in}: as one moves a large distance in field space, a tower of states becomes exponentially light leading to a dual description. Its extension to the space of metrics, the AdS Distance Conjecture~\cite{Lust:2019zwm}, states that as the cosmological constant is taken to zero, a tower of states emerges with mass scale
\begin{equation}
m \sim \lambda\, |\Lambda|^{\alpha}\,,
\label{eq:adc}
\end{equation}
in Planck units, with $\alpha$ an order-one positive constant.  There is strong evidence for this conjecture for $\Lambda<0$ studied in the holographic context, and it is natural to expect that it will also apply to $\Lambda>0$. Applying this to the observed dark energy \eqref{eq:cc}, and using the general expectation $\tfrac1d\le \alpha\le \tfrac12$ for the exponent together with experimental constraints, fixes the exponent: the tower cannot be much heavier than $\Lambda^{1/4}$, while $\alpha>1/4$ would give a tower so light as to conflict with the lack of deviation of Newton's inverse square law up to $30\mu$ m~\cite{Lee:2020zjt}. One is led to~\cite{Montero:2022prj}\footnote{Note that the distance conjecture leads to a solution of cosmological constant problem, namely if it is small it is pegged to the light tower mass $\Lambda \sim m_{light}^4$ and not the UV scale.}
\begin{equation}
\mKK \sim \lambda^{-1}\,\Lambda^{1/4}\,,\qquad 10^{-4}\lesssim\lambda\lesssim 10^{-1}\,,
\end{equation}
i.e., a KK tower in the $10^{-1}\,$--$\,10\,$meV range, corresponding to \emph{one} extra dimension of size
\begin{equation}
\ell_5 \sim \lambda\Lambda^{-1/4}\,\sim\, 0.1\,\text{--}\,10~\mu\text{m}\,.
\label{eq:l5}
\end{equation}
This is consistent with the current experimental tests of Newton's inverse-square law, which is verified down to distances of roughly $30\,\mu$m~\cite{Lee:2020zjt}.  The associated five-dimensional Planck scale is
\begin{equation}
M_5 \sim \mKK^{1/3}\,\Mpl^{2/3}\,\sim \Lambda^{\frac{1}{12}}\sim\, 10^{9}\,\text{--}\,10^{10}~{\rm GeV}\,,
\label{eq:M5}
\end{equation}
which will play the role of a fundamental UV scale in what follows.

Why one dimension, and why a KK tower rather than a string tower? By the Emergent String Conjecture~\cite{Lee:2019wij}, the light tower at infinite distance is either a KK tower or a tower of string excitations. A light string tower at the meV scale is excluded: the string scale would lie far below collider energies, in conflict with the LHC. Similarly, $n\ge 2$ large extra dimensions decompactifying uniformly at the dark energy scale would give a higher-dimensional Planck scale
\begin{equation}
n\ge2:\qquad \widehat M_{\rm pl}\lesssim 10~{\rm TeV}\,,
\end{equation}
which is essentially excluded by the LHC\footnote{The fine-tuned possibility with two dark dimensions of micron size has been suggested in~\cite{Anchordoqui_2025} but faces difficulties involving fine-tuning initial temperature very close to that needed for BBN~\cite{HallSmith1999} (as not to overproduce dark matter), as well as the astrophysical constraints from SN1987A cooling by KK emission which further push toward the lower end of this window, $\ell_5\lesssim 0.2\,\mu$m~\cite{Hardy:2025hkt}.  Also the axion physics and neutrino physics discussed in this paper do not work well for 2 extra dimensions.}.

\begin{figure}[t]
\centering
\includegraphics[width=0.72\textwidth]{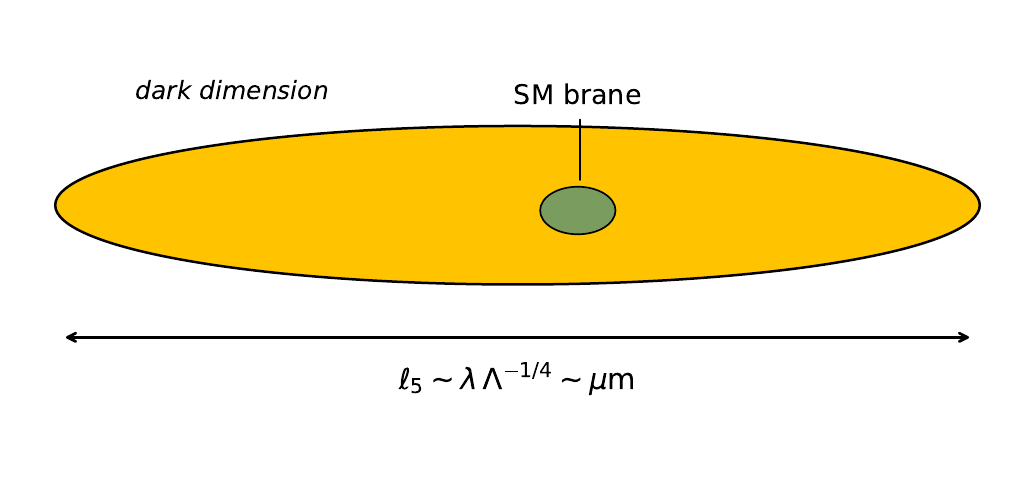}
\caption{The Standard Model is localized on a brane in the dark dimension, whose size is fixed by the dark energy, $\ell_5\sim\lambda\Lambda^{-1/4}\sim\mu$m. Only gravity and neutral bulk fields propagate in the fifth dimension.}
\label{fig:brane}
\end{figure}

The SM must be localized on a brane in the dark dimension: otherwise every SM field would come with a KK tower with spacing in the meV--eV range, which is grossly excluded. The picture is that of a codimension-one brane in a five-dimensional bulk, as in Figure~\ref{fig:brane}. Only gravity (and possibly other neutral bulk fields, such as a $B-L$ gauge field and right-handed neutrinos, see Section~\ref{sec:neutrinos}) propagates in the dark dimension.

\section{Dark Matter: the Kaluza--Klein Graviton Tower}
\label{sec:dm}

The dark dimension comes with a natural dark matter candidate. The distance conjecture predicts that there should be a light weakly coupled tower, which in our case is the tower of KK excitations of the graviton~\cite{Gonzalo:2022jac,Law-Smith:2023czn,Obied:2023clp}.\footnote{See~\cite{anchordoqui2024landscapeswamplandextradimensions} and references therein for an alternative dark matter scenario in the dark dimension involving primordial black holes.} These dark gravitons couple to the SM only gravitationally, and within the tower they cascade\footnote{The cascade can proceed in a way consistent with cosmological
   observations unless one restricts the action for the fields in the internal
   dimensions to two-derivative terms, a restriction that would rule out not
   only dS solutions \cite{Maldacena:2000mw}, including unstable or quasi-dS
   ones, but also potentially a cascade consistent with cosmology
   \cite{Lee:2026kkg,Langhoff:2026wiz}. This can be avoided by including
   suitable higher derivative terms, or by the conversion of the KK graviton
   tower into another KK tower. As discussed in Appendix~\ref{app:phitower},
   the evolving dark sector of Section~\ref{sec:fds} supplies a natural
  candidate for the latter: the KK tower of the quintessence field itself.} by decaying preferentially to lower KK levels (``dark-to-dark'' decays), so that the tower behaves as slowly decaying dark matter with an evolving mass distribution around $\mKK\sim$ meV--eV, with a small loss of mass for each decay converted to kick velocity of the order of $v_{kick}\sim1-10$ km/s.

\begin{figure}[t]
\centering
\includegraphics[width=0.72\textwidth]{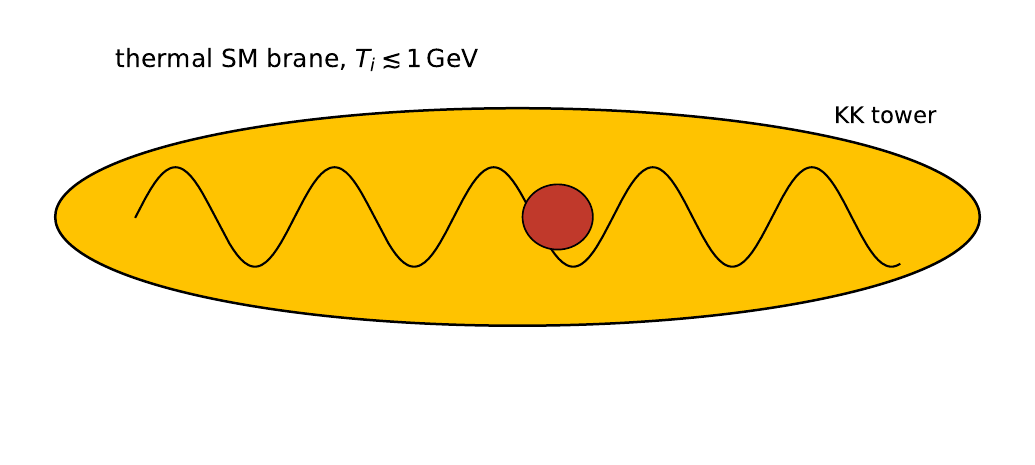}
\caption{Dark gravitons: the thermal SM brane, at temperatures $T\lesssim T_i\sim1$~GeV, radiates KK gravitons into the dark dimension. The resulting tower of dark gravitons constitutes the dark matter.}
\label{fig:darkgraviton}
\end{figure}

How is this dark matter generated? Cosmology provides the answer: the SM brane is in thermal equilibrium, but the bulk tower is not. It is natural to expect the tower to be generated by the five-dimensional gravitational coupling of the bulk gravitons to the thermal energy--momentum on the brane, as in Figure~\ref{fig:darkgraviton}. The production is dominated by the highest temperatures, and requiring the resulting abundance to match the observed dark matter density fixes the maximal (``initial'') temperature of the radiation era to be
\begin{equation}
T_i\,\sim\, 1~{\rm GeV}\,,
\label{eq:Ti}
\end{equation}
with $T_i$ bounded below by BBN, $T_{\rm BBN}\sim 1\ {\rm MeV} \lesssim T_i$. Conversely, if the radiation era begins at $T_i\sim1$~GeV, one \emph{automatically} obtains the right amount of dark matter. In the dark-dimension framework, the origin of the initial temperature is tied to the stabilization of the internal geometry. Let $m_{\rm mod}$ denote the smallest mass among the moduli that fix the compact dimensions and the position, width, and shape of the Standard Model brane. Before these fields settle into their minima, a reliable four-dimensional cosmology cannot be defined. Since their couplings are generically Planck suppressed, their decay rate scales as $\Gamma_{\rm mod}\sim m_{\rm mod}^3/M_{\rm Pl}^{2}$. The TCC implies that a quasi-de Sitter phase cannot persist parametrically longer than a Hubble time, up to logarithmic corrections, so the moduli must decay and stabilize the compactification within this time: $\Gamma_{\rm mod}\gtrsim H_\Lambda$, where $H_\Lambda\sim(\Lambda)^{1/2}$. Combining these relations gives (in Planck units)
$$m_{\rm mod}\gtrsim {\Lambda}^{\frac{1}{6}},$$
 parametrically of order the GeV scale. It is therefore natural for the radiation-dominated era to begin only after the moduli have decayed, at a temperature $T_i<m_{\rm mod}$. The thermal production of Kaluza--Klein gravitons is dominated by this highest radiation temperature, and taking $T_i\sim\Lambda^{1/6}M_{\rm Pl}^{1/3}\sim{\cal O}(1){\rm GeV}$ naturally yields the observed dark-matter abundance. The same scale may also underlie the electroweak hierarchy: if the Higgs sector is controlled by the moduli fixing the Standard Model brane, one is led parametrically to $m_{\rm weak}\sim\Lambda^{1/6}$ in Planck units. The mass of the modulus is naturally related both to the size of the SM brane as well as the mass scale on the brane, namely the weak scale. Thus the TCC connects moduli stabilization, the onset of radiation domination, the weak scale, and the dark-matter abundance, turning the latter from an independent input into an output of the cosmological framework.

\section{Neutrinos and $B-L$}
\label{sec:neutrinos}

One of the striking numerical coincidences in nature is that between the neutrino mass scale and the dark energy scale, $m_\nu \sim \Lambda^{1/4}$. The dark dimension offers a natural explanation~\cite{Montero:2025rvz}.

The symmetry $B-L$ is a global symmetry of the SM (non-anomalous gauge symmetry only if right-handed neutrinos exist). Quantum gravity does not admit exact global symmetries, so one expects $B-L$ to be gauged. In the dark dimension setup this can happen in two ways. Either as part of the broken GUT gauge group on the SM brane or on the bulk.  In the latter case a $B-L$ gauge field propagating in the five-dimensional bulk induces what looks like a global symmetry on the SM brane. Its gauge anomaly is canceled by charged bulk modes, and this \emph{requires} the existence of three right-handed neutrinos $\nu_R$ propagating in the dark dimension, which couple to the brane-localized lepton doublets through the Higgs. The five-dimensional wavefunction spreading of the bulk $\nu_R$ then suppresses the Dirac masses down to the KK scale, leading to the natural matching
\begin{equation}
m_\nu \,\sim\, \mKK\,\sim\,\Lambda^{1/4}\,\sim\, {\rm meV}\text{--}0.1~{\rm eV}\,,
\label{eq:mnu}
\end{equation}
thereby explaining the observed coincidence between neutrino masses and the dark energy scale. Higgsing the bulk $B-L$ gauge symmetry by a bulk scalar at the Higgs scale results in a massive gauge field with
\begin{equation}
m_{B-L}\gtrsim 100~{\rm GeV}\,,\qquad g_{B-L}\sim 10^{-10}\,,
\end{equation}
which is allowed by current bounds and potentially accessible to future experiments. The scenario moreover predicts a tower of massive sterile right-handed neutrinos, driven by the B-L gauge symmetry breaking and in addition the mass of the active neutrinos come out in the right range
\begin{equation}
m_\nu^{\rm active}\sim {y^2H^2\over M_5}\sim {\Lambda^{2/6}\over \Lambda^{1/12}}\sim \Lambda^{1/4}\sim {\rm meV}\ ,
\end{equation}
where $y\sim 10^{-3}$ is a Yukawa coupling between bulk and brane fermions.

\section{The QCD Axion and the Weak Gravity Conjecture}
\label{sec:axion}

Given that QCD is localized on the SM brane, it is natural to assume that the QCD axion, if it exists, is also localized on the SM brane~\cite{Gendler:2024adn}. For a brane-localized axion the relevant gravitational cutoff is not the four-dimensional Planck scale but the five-dimensional one, $M_5\sim 10^{9}$--$10^{10}$~GeV of Eq.~\eqref{eq:M5}. The axionic version of the Weak Gravity Conjecture~\cite{Arkani-Hamed:2006emk} then bounds the axion decay constant by this scale:
\begin{equation}
f_a\,\lesssim\, M_5\,\sim\,10^{9}\text{--}10^{10}~{\rm GeV}\,.
\label{eq:fa}
\end{equation}
But this is also the experimental lower bound on $f_a$!  So this fixes $f_a\sim 10^{9}\text{--}10^{10}~{\rm GeV} $.
This leads to two predictions: (1) the QCD axion decay constant lies at (or just below) the five-dimensional Planck scale, so that the axion mass falls in the range
\begin{equation}
m_a\,\sim\,\frac{m_\pi f_\pi}{f_a}\,\gtrsim\,0.1\text{--}1~{\rm meV}\,,
\end{equation}
once again the same meV scale that controls the dark sector; and (2) axions in this window are within reach of the next generation of axion experiments (e.g.\ helioscopes such as IAXO and dedicated searches in the meV band). It is amusing that the dark dimension independently points to the classic astrophysical window for $f_a$.

\section{Grand Unification and the Dark Dimension}
\label{sec:gut}

At first sight a fundamental scale $M_5\sim10^{9}$--$10^{10}$~GeV seems to leave no room for conventional grand unification at $M_{\rm GUT}\sim10^{16}$~GeV. In fact the opposite is true: the dark dimension naturally \emph{relates} the GUT scale to the weak scale~\cite{Heckman:2024oot}.

\begin{figure}[t]
\centering
\includegraphics[width=1.1\textwidth]{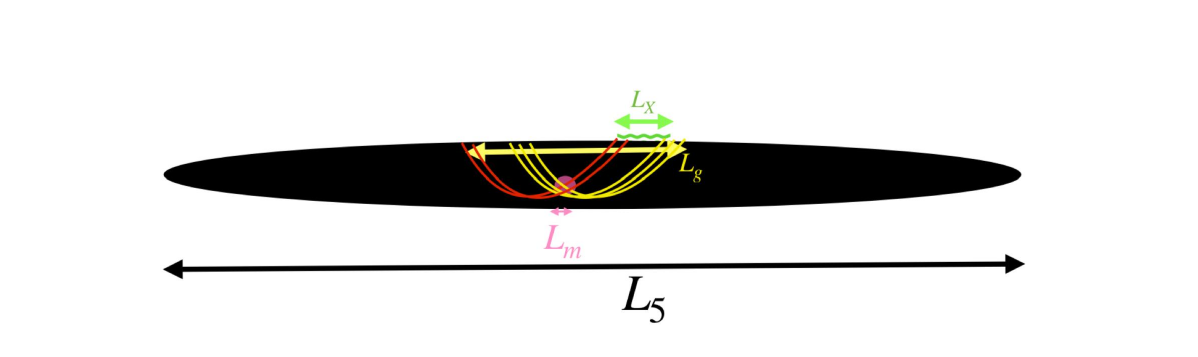}
\caption{The SM brane wraps a small internal cycle of size $L_X$ inside the compactification, within the dark dimension of size $L_5$. GUT-scale $X$ bosons arise as winding modes, $m_X\sim M_5^2 L_X$, while momentum modes $\sim L_X^{-1}$ lie not far above the weak scale, leading to $M_{\rm GUT}M_{\rm weak}\sim M_5^2$.}
\label{fig:gut}
\end{figure}

The idea is that the SM gauge fields arise on a brane which wraps a small internal cycle, as in Figure~\ref{fig:gut}. The heavy $X$ gauge bosons responsible for proton decay arise as \emph{winding} modes on this cycle, while the light charged states arise as \emph{momentum} modes. With brane tension set by the fundamental scale, $T\sim M_5^2$, a cycle of size $L_X$ gives
\begin{equation}
m_X\,\sim\, T\, L_X\, \sim\, M_5^2\, L_X\,\gtrsim\,10^{16}~{\rm GeV}\,,
\qquad
L_X^{-1}\,\gtrsim\,10^{4}~{\rm GeV}\,,
\label{eq:gutrel}
\end{equation}
so that GUT-scale winding modes coexist with momentum modes not far above the weak scale. Eliminating $L_X$, the two scales are tied to the fundamental scale by
\begin{equation}
M_{\rm GUT}\cdot M_{\rm weak}\,\sim\, M_5^{2}\,,
\label{eq:gutweak}
\end{equation}
which is beautifully consistent: $\sqrt{10^{16}\times 10^{2\text{--}4}}\sim10^{9}$--$10^{10}$~GeV~$\sim M_5$. In this picture the states mediating proton decay are string-like ($m_X\gtrsim10^{16}$~GeV corresponds to a winding string with tension $T\sim M_5^2$), and the proximity of the weak scale to $L_X^{-1}$ links the hierarchy problem to the geometry of the internal cycle.  On the other hand as we noted above the moduli fields fixing the SM brane should have mass above GeV which is consistent with them leading to a stabilized moduli with $L_X^{-1}\sim 10\ \rm  TeV$. The hierarchy between the weak and GUT scales is thus reinterpreted as a modest hierarchy of brane geometry inside the dark dimension setup.

We thus see that the weak scale $10^2 $ GeV could be related not only to $\Lambda^{1/6}\sim 1$ GeV by TCC, but also to inverse of SM brane size ${\sim 10^4 \ \rm GeV}$, by a mild hierarchy of $10^{\pm 2}$.

\section{The Evolving Dark Dimension and the DESI Results}
\label{sec:fds}

The Swampland de Sitter conjecture~\cite{Obied:2018sgi} and the TCC~\cite{Bedroya:2019snp} suggest that dark energy is not a cosmological constant: the potential of the field $\phi$ controlling it cannot remain flat over long stretches of field space. In its refined form, the TCC implies two constraints in Planck units. First, over any large monotonic stretch of field space the potential must fall off at least as fast as $\exp\!\big(-2\,\Delta\phi/\sqrt{(d-1)(d-2)}\big)$, i.e.\ the \emph{average} slope obeys\footnote{Note that throughout the paper we are setting the 4d Planck mass to 1.}
\begin{equation}
\left\langle \frac{|\nabla V|}{V}\right\rangle \,\ge\, \frac{2}{\sqrt{(d-1)(d-2)}}\;\overset{d=4}{=}\;\sqrt{\tfrac23}\,\simeq\,0.82\,.
\label{eq:tccavg}
\end{equation}
Second, in the asymptotic regions of field space the slope is bounded below by~\cite{Bedroya:2019snp} (see also~\cite{Rudelius2019})
\begin{equation}
\frac{|\nabla V|}{V}\,\ge\,\frac{2}{\sqrt{d-2}}\;\overset{d=4}{=}\;\sqrt{2}\, ,
\label{eq:tccasym}
\end{equation}
which is consistent with the known asymptotic limits of string theory, decompactification or emergent string limits~\cite{Lee:2019wij}.

Motivated by these ideas and the fact that dark matter mass also naturally has logarithmic derivative with respect to moduli of $O(1)$ in Planck units led to the cosmological model of fading dark sector \cite{Agrawal:2019dlm} given by two parameters of $O(1)$ in Planck units $(c,c')$ fixing the logarithmic slopes of $V(\phi),m(\phi)$ which can be assumed to be approximately constant and thus taken to be
\begin{equation}
V(\phi)=V_0\,e^{-c\,\phi}\,,\qquad
\mDM(\phi)=m_0\,e^{-c'\phi}\,,
\label{eq:fds}
\end{equation}
One should bear in mind that the prediction of order 1 numbers $c,c'$ is a rather non-trivial prediction, as for example it predicts $|\nabla V|\sim 10^{-122\pm 1}$ in Planck units!

The dark dimension provides a concrete realization of such an evolving dark sector~\cite{Vafa:2024fpx,Bedroya:2025fwh}: the field $\phi$ controlling the dark energy has a component along the radion of the dark dimension itself, so that dark energy evolves through the evolution of the dark dimension. As $\phi$ evolves, the radius of the dark dimension varies, and therefore so does the mass of the dark matter, which consists of its KK gravitons. The evolution of dark energy and of the dark matter mass are thus \emph{correlated}, as both descend from the same scalar field.  The energy fades in each epoch namely, in the matter dominated era, dark matter goes down and in the energy dominated era, dark energy goes down, which is why this model is called the ``fading dark sector'' (FDS).  It is adequately captured by the
two-parameter cosmological model given by $(c,c')$. As we will now review the data turn out to prefer $c<0,\ c'>0$, corresponding to a decrease in mass of dark matter in the earlier era (expanding of dark dimension) and increasing of the mass in the energy dominated era (shrinking dark dimension).  Dark energy variation is opposite in direction to that of dark matter.  This variation of dark matter mass produces an apparent phantom equation of state $w<-1$ when the data are analyzed assuming standard dark matter with non-varying mass.

\begin{figure}[!htbp]
    \centering
    \includegraphics[width=1.1\linewidth]{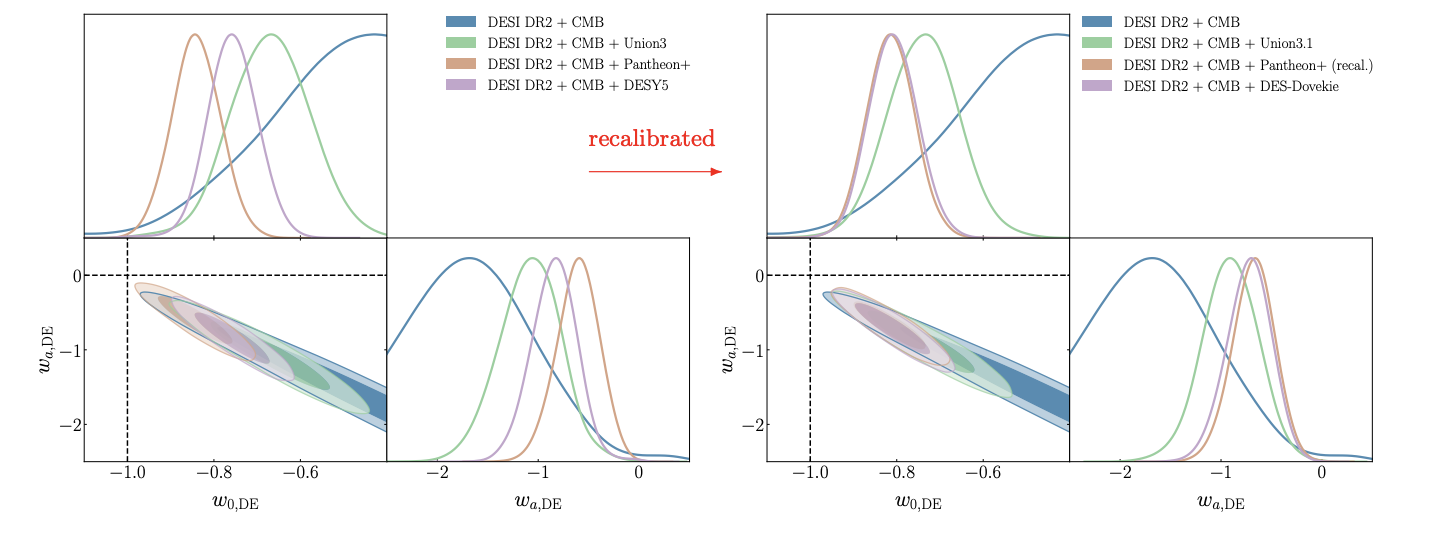}
    \includegraphics[width=1.1\linewidth]{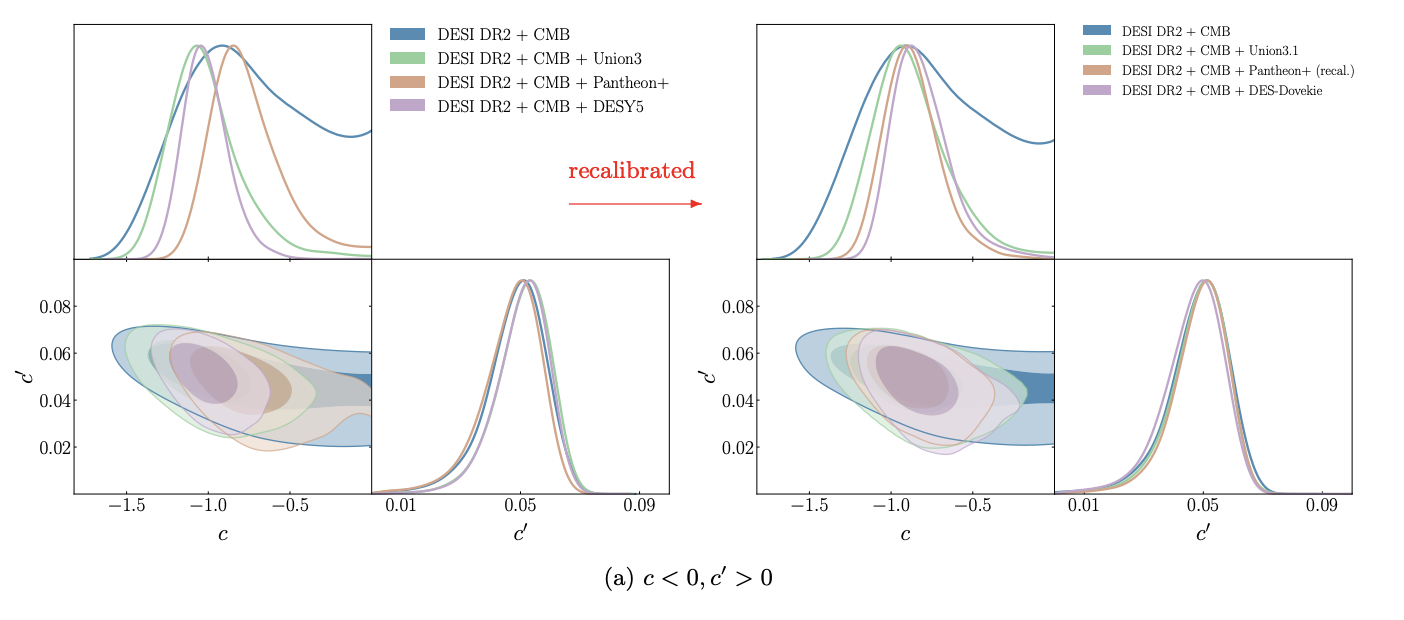}
    \caption{Constraints from DESI DR2 + CMB combined with supernova datasets~\cite{Bedroya:2025fwh}. Top: the $w_0w_a$CDM (CPL) parametrization; bottom: the FDS model parameters $(c,c')$. In each row, the left panels use the original supernova samples (Union3, Pantheon$+$, DESY5) and the right panels the recalibrated ones (Union3.1, Pantheon$+$ recal., DES-Dovekie). The preferred FDS region has $c\simeq-1$, $c'\simeq0.05$, stable across dataset combinations.}
    \label{fig:fdsfit}
\end{figure}

MCMC analyses for FDS using DESI DR2 combined with CMB and various supernova datasets (DES, Pantheon$+$, Union) were performed in~\cite{Bedroya:2025fwh}. The results are summarized in Table~\ref{tab:fds} and Figure~\ref{fig:fdsfit}. Two features stand out. First, the FDS model matches the statistical performance of the phenomenological $w_0w_a$CDM (CPL) parametrization, with preference over $\Lambda$CDM ranging from $2.5\sigma$ to $4.0\sigma$ depending on the supernova dataset, but unlike $w_0w_a$CDM, it does so without invoking an actual phantom crossing, with a microscopic interpretation, and with only two parameters directly tied to the dark dimension. Second, the preferred parameters are remarkably stable across datasets and indeed $O(1)$ in Planck units:
\begin{table}[h]
\centering
\small
\begin{tabular}{lcccc}
\toprule
& \multicolumn{2}{c}{$w_0w_a$CDM} & \multicolumn{2}{c}{FDS $(c<0,\ c'>0)$}\\
\cmidrule(lr){2-3}\cmidrule(lr){4-5}
Datasets & $\sigma$ & $\sigma$ & $c$ & $c'$ \\
\midrule
DESI+CMB                       & $3.0$ &  $2.5$ & $-0.72^{+0.45}_{-0.43}$ & $0.05\pm0.01$ \\
DESI+CMB+Union3               & $3.7$ &  $3.4$ & $-1.01^{+0.21}_{-0.22}$ & $0.05\pm0.01$ \\
DESI+CMB+Union3.1                 & $3.4$ &  $3.0$ & $-0.87\pm0.24$          & $0.05\pm0.01$ \\
DESI+CMB+Pantheon$+$          & $2.7$ &  $2.9$ & $-0.76^{+0.23}_{-0.22}$ & $0.05\pm0.01$ \\
DESI+CMB+Pantheon$+$ (recal.)   & $3.2$ &  $3.2$ & $-0.87\pm0.18$          & $0.05\pm0.01$ \\
DESI+CMB+DESY5                 & $4.1$ &  $4.0$ & $-1.00^{+0.13}_{-0.14}$ & $0.05\pm0.01$ \\
DESI+CMB+DES-Dovekie             & $3.1$ &  $3.2$ & $-0.81\pm0.19$          & $0.05\pm0.01$ \\
\bottomrule
\end{tabular}
\caption{Fits of the fading dark sector model \eqref{eq:fds} to DESI DR2 + CMB combined with various supernova datasets, compared to the $w_0w_a$CDM (CPL) parametrization~\cite{Bedroya:2025fwh}. Shown are the improvement relative to $\Lambda$CDM (where available), the corresponding significance, and the marginalized constraints on $(c,c')$. Rows labeled ``recal.''\ and the Union3.1/DES-Dovekie rows use recalibrated supernova samples. The best-fit (MAP) values of $c$ cluster around $c\simeq-1$.}
\label{tab:fds}
\end{table}

\begin{figure}[!htbp]
\centering
\includegraphics[width=0.6\textwidth]{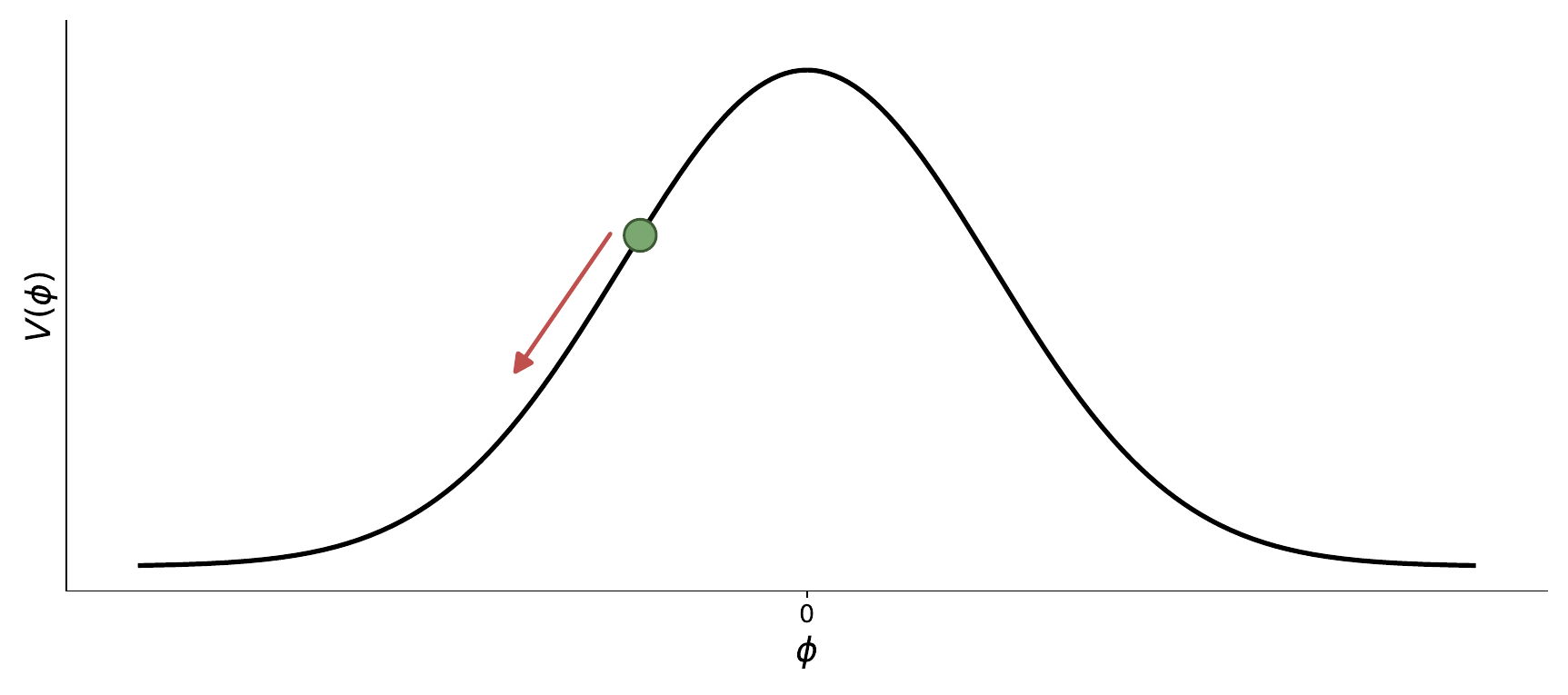}
\caption{The fitted slope $|c|\simeq1\pm0.2$, being smaller than the asymptotic value $\sqrt2$ required by the TCC together with string-theoretic asymptotics, suggests that the dark energy field lies in the interior of moduli space, near an unstable critical point of the potential.}
\label{fig:hilltop}
\end{figure}

It is remarkable that both $c,c'$ are consistent across all the datasets.  The fact that $c'\sim .05\sim O(1)$ is as expected specially if one recalls there is a cosine projection factor needed for the component of scalar along the radion of the dark dimension.\footnote{An alternative realization can arise if the graviton KK tower is converted to a KK tower of $\phi$.  In this case it is possible to realize the fading dark sector scenario with a fixed size of the dark dimension: $c'$
can arise instead from the dependence on $\phi$ of the propagation speed of its
KK tower, as discussed in Appendix~\ref{app:phitower}.} 
Exchange of the $\phi$ field, which is nearly massless, leads to a long range attractive 5-th force which acts only on the dark matter and its strength is $2c'^2\sim0.5\%$ of the gravitational attraction. Indeed, this is consistent with the current upper bound on the long range 5-th force in the dark sector~\cite{Kesden:2006zb}
which leads to $c'<0.2$.

It is reassuring that $|c|\simeq 1\pm0.2$: it is of order one and in the range expected for the \emph{average} slope, Eq.~\eqref{eq:tccavg}. At the same time, the fitted slope is \emph{smaller} than the minimum value $\sqrt2\simeq1.4$ that it should be above asymptotically, Eq.~\eqref{eq:tccasym}. This means the field $\phi$ is not in an asymptotic regime: it must be in the \emph{interior} of moduli space, perhaps close to an (unstable) critical point of the potential, as sketched in Figure~\ref{fig:hilltop}. This raises a sharp question for string theory:
\begin{quote}
\emph{How do we construct examples in string theory of points in the interior of field space where $V>0$ with small slope---in particular, (unstable) critical points without supersymmetry?}
\end{quote}
This question motivates the second part of this paper.

\section{Hierarchies of Scales as Powers of $\Lambda$}
\label{sec:hierarchy}

Dirac famously asked why nature presents us with such strange small (and large) numbers. The dark dimension suggests an updated version of the answer: the mysterious scales of our universe organize themselves as simple rational powers of the dark energy, in Planck units:
\begin{equation}
\begin{aligned}
\Lambda
&\sim 10^{-120}\,, \\[1mm]
H_0\sim\tau_{\rm now}^{-1}
&\sim \Lambda^{1/2}
\sim 10^{-60}
\sim 10^{-41}~{\rm GeV}\,, \\[1mm]
m_\nu\,,\ m_a\,,\ \mKK
&\sim \Lambda^{1/4}
\sim 10^{-30}
\sim 10^{-12}\text{--}10^{-10}~{\rm GeV}\,, \\[1mm]
\Lambda_{\rm QCD}\sim\alpha\,\Lambda_{\rm weak}
&\sim \Lambda^{1/6}
\sim 1~{\rm GeV}\,, \\[1mm]
M_{\rm weak}\,M_{\rm GUT}
&\sim \left(\Lambda^{1/12}\right)^2
\sim M_5^2
\sim \left(10^{9\text{--}10}~{\rm GeV}\right)^2\,.
\end{aligned}
\label{eq:ladder}
\end{equation}
The first two lines are standard; the third is the content of Sections~\ref{sec:dd}--\ref{sec:axion}; the fourth line encodes the intriguing observation that the QCD scale sits at roughly $\Lambda^{1/6}$ which is natural for the moduli mass of SM brane based on TCC, one gauge-coupling factor below the weak scale; and the last line is the GUT relation \eqref{eq:gutweak} of Section~\ref{sec:gut}. What would have appeared as a random scatter of hierarchies becomes a ladder of powers $\Lambda^{1/2},\Lambda^{1/4},\Lambda^{1/6},\Lambda^{1/12}$ of a single small quantity---whose smallness, in turn, the Swampland program ties to the existence of the dark dimension.

\section{Non-Supersymmetric Vacua: Generalities}
\label{sec:nonsusy}

We now turn to the second part of this paper. There is no supersymmetry at low energies in our universe. But non-supersymmetric string vacua are not well understood: almost all reliable constructions of the string landscape lean on spacetime supersymmetry. Without supersymmetry we expect a nonzero potential for the light moduli $\phi$; and indeed our universe has
\begin{equation}
V\,\sim\,10^{-122}\,\Mpl^4\,\neq\,0\,,
\end{equation}
motivating quintessence-like rolling, $V\to0$, in asymptotic or interior regions of field space, as discussed in Section~\ref{sec:fds}.

The theoretical prejudice of the 1990s was that, despite the fact that lack of supersymmetry predicts otherwise, the vacuum energy is exactly zero because experimental bound gave $V< 10^{-120}$: the observed small positive $V\sim 10^{-122}>0$ came as a surprise!  Similarly one may have assumed since experimentally we know $|\nabla V|< 10^{-120}$ it should be $\nabla V=0 $ as in $\Lambda CDM$, but as we have seen the Swampland ideas predict otherwise $|\nabla V|\sim V\sim 10^{-122}$ and seem to be fitting current data better than $\Lambda CDM$!
The data-driven lesson of Section~\ref{sec:fds}---that we appear to sit in the interior of moduli space near a positive-energy critical point---turns the construction of non-supersymmetric critical points from a theoretical curiosity into a phenomenological priority. The main difficulty is the absence of the usual supersymmetric protection: generic loop corrections displace candidate critical points. The strategy that has proven effective is to use \emph{symmetries}, and in particular unbroken \emph{duality} symmetries: isolated points in field space fixed by a symmetry are automatically critical points of any invariant potential, to all orders.

\section{Isolated Critical Points from Symmetries: Non-Geometric Vacua}
\label{sec:symmetries}

\noindent{\bf Strategy:} \emph{find isolated points in the light moduli space which are fixed by discrete gauge symmetries.} At such points $dV=0$ is guaranteed by symmetry, regardless of the detailed form of quantum corrections.
This strategy has been used in the early days of studying non-supersymmetric critical points of $V$ \cite{Ginsparg:1986wr}. 

More recently non-geometric flux compactifications~\cite{Becker:2006ks} have been revisited. Mirror duals of rigid Calabi--Yau manifolds with flux described by Landau--Ginzburg (LG) orientifold models such as the $(x^3)^{\otimes 9}$ and $(x^4)^{\otimes 6}$ models are a useful arena to find stabilized four-dimensional vacua with no geometric moduli. Recently these constructions have been revisited ~\cite{Becker:2024ijy,Rajaguru:2024emw}. In~\cite{Chen:2025rkb} this class of vacua has been further studied and symmetry-based arguments were given for the existence of these isolated dS vacua: the unbroken duality symmetries of the LG models fix isolated points of the complex moduli space and symmetric points of axio-dilaton\footnote{Symmetries can generally be used to argue for critical points for modular superpotentials \cite{Mohseni:2025tig}.}, protect the critical points beyond perturbative reasoning. In particular it is found that
\begin{equation}
V>0, \quad \nabla V=0:\quad\text{(unstable) de Sitter critical points}\,,
\end{equation}
i.e.\ dS saddles of the type suggested by the cosmological considerations of Section~\ref{sec:fds}. Other isolated dS critical points, which are not under full theoretical control have also been studied recently (by considering Casimir energies on Riemann-flat manifolds in M-theory) in~\cite{BentoMontero:2025}.

\section{Scherk--Schwarz Compactifications of M-theory}
\label{sec:ss}

The main challenge in extending this strategy is to identify the \emph{duality groups of non-supersymmetric vacua}. A clean laboratory is provided by Scherk--Schwarz compactifications of M-theory~\cite{Scherk:1979zr,Baykara:2026scherk}: toroidal compactifications with a non-trivial spin structure, i.e.\ M-theory on $T^n$ with anti-periodic boundary conditions for fermions along (at least) one circle. Supersymmetry is broken at the compactification scale, and a one-loop Casimir potential is generated on the moduli space of flat metrics and $C$-field holonomies.

The key point of~\cite{Baykara:2026scherk} is that a subgroup of the U-duality group survives the Scherk--Schwarz boundary conditions: the unbroken duality group is the subgroup preserving the spin structure (roughly, a congruence-type subgroup of $E_{n(n)}(\mathbb{Z})$). One can then find isolated points of the moduli space fixed by elements of this unbroken duality group; by the symmetry argument these are exact critical points of the potential. Concretely, one finds critical points of the Casimir potential for M-theory compactified on
\begin{quote}
\emph{a Planck-size cubic torus supplemented by half-integral $C$-field holonomy}\,,
\end{quote}
and this construction extends to compactifications down to $d=8,6,5$ and $4$ spacetime dimensions~\cite{Baykara:2026scherk}. Notably, the critical points sit at Planckian size---deep in the interior of moduli space, precisely the regime suggested by the cosmological fits of Section~\ref{sec:fds}---so the quantitative features of the potential there are not under full control; but their existence as critical points is protected by the unbroken duality symmetry.  Unfortunately one cannot determine the sign of $V$ in these cases.\footnote{An important structural byproduct is a duality-covariant consistency condition: for M-theory on $T^n$ orbifolded by discrete $\mathbb{Z}_N$ symmetries acting as phases on the charge lattice, a duality-invariant anomaly cancellation condition was conjectured in~\cite{Baykara:2026scherk} which generalizes the level-matching requirement of perturbative string orbifolds. For compactifications where a perturbative string description exists, the condition reduces to standard level matching; in general it is a genuinely U-duality invariant statement about the charge lattice, providing a consistency check on non-supersymmetric M-theory backgrounds that have no weakly coupled corner.}

The exploration of dualities for non-supersymmetric theories have been further taken up in the recent work leading to a \emph{duality web for non-supersymmetric strings}~\cite{Baykara:2026web}, which includes, remarkably, further evidence for dualities between the bosonic string and superstrings as was anticipated in \cite{Bergman:1997rf,Dudas:2025ubq}. See also the related works~\cite{Altavista:2026horava,Altavista:2026ouroboros}, and other recent works~\cite{Dasgupta:2026wedge,Basile:2026wedge}, for closely related developments. Non-supersymmetric string vacua are beginning to get demystified.

\section{Conclusions}
\label{sec:conclusions}

Swampland principles lead to a unique corner of the quantum gravity landscape---the dark dimension---which beautifully fits the various experimental bounds and leads to new predictions, which so far match the data remarkably well: a micron-range extra dimension potentially visible in next-generation tests of Newton's law; KK graviton dark matter with the right abundance for $T_i\sim1$~GeV; neutrino masses tied to $\Lambda^{1/4}$ with a weakly coupled $B-L$ gauge boson and sterile neutrinos\footnote{The sterile neutrino masses should be above $\sim 100$ GeV to avoid issues with cosmology \cite{Anchordoqui:2026ret}.}; an axion decay constant at $M_5\sim10^{9\text{--}10}$~GeV; a GUT--weak relation $M_{\rm GUT}M_{\rm weak}\sim M_5^2$; and an evolving dark sector which accounts for the DESI/DES hints of phantom dark energy with two microscopic parameters, $c\simeq-1$ and $c'\simeq0.05$. We may be on the verge of verifying predictions from string theory for the first time.

At the same time, the same fits tell us that we live in the interior of moduli space, near a positive-energy critical point, in a universe with no low-energy supersymmetry. Meeting this challenge requires developing string theory away from its supersymmetric comfort zone. The symmetry-based strategy reviewed in the second part of this paper---isolated critical points protected by unbroken duality groups, realized both in non-geometric flux vacua and in Scherk--Schwarz compactifications of M-theory, together with the emerging duality web for non-supersymmetric strings---suggests that this challenge is beginning to be met. It is tempting to speculate that the two halves of this story will merge: that the critical point our universe appears to be near will ultimately be understood as one of the symmetry-protected non-supersymmetric critical points of string theory.

\section*{Acknowledgments}

I would like to thank my collaborators on the works reviewed here: Luis Anchordoqui, Alek Bedroya, Zihni Kaan Baykara, Shi Chen, Matilda Delgado, Emilian Dudas, Cora Dvorkin, Naomi Gendler, Eduardo Gonzalo, Damian van de Heisteeg, Jonathan Heckman, Dieter L\"ust, Miguel Montero, Georges Obied, Eran Palti, Hector Parra De Freitas, Irene Valenzuela, Timo Weigand, David Wu and Fengjun Xu. I also thank the organizers of String Phenomenology 2026 in Paris and of Strings 2026 in Shanghai for putting together two stimulating conferences. 

I used Claude (Anthropic) for editorial assistance and as a sounding board in preparing this manuscript. This work is supported in part by a grant from the Simons Foundation (602883, CV), the DellaPietra Foundation, and by the NSF grant PHY-2013858.  

\begin{appendices}\section{KK tower of $\phi$ as Dark Matter and FDS with a stabilized Dark Dimension}
\label{app:phitower}

In this appendix we present two variations of the FDS cosmology discussed in Section~\ref{sec:fds}. One is dark matter as the KK tower of the $\phi$ field instead of the gravitons. The other, which is natural to consider if the KK tower of $\phi$ is the dark matter, is a stabilized dark dimension, which still leads to a changing mass scale for the KK tower of $\phi$, as we discuss below.

The fits of Section~\ref{sec:fds} contain a piece of information about the five-dimensional theory that bears on the fate of the KK gravitons discussed in the footnote of Section~\ref{sec:dm}. The quintessence field $\phi$ of \eqref{eq:fds} is light enough to roll today, so if it is a 5-dimensional field in the dark dimension scenario, then it has to be essentially massless in 5d. Indeed we now argue that $\phi$ must be a 5d massless field: its zero mode moves the radius of the dark dimension, but only at the level $c'\simeq0.05$, against the value $\sqrt\frac{(D-2)}{(d-2)(D-d)}=\sqrt{\frac{3}{2}}$ it would have if it were purely the 4d radion field. As noted in \cite{Bedroya:2025fwh}, $\phi$ is not the radion but has a small projection on it, $\sin\theta=\frac{c'}{\sqrt{3/2}}\simeq0.04$: it is mostly a modulus of the smaller internal dimensions. A modulus of the internal geometry is a 5d bulk field, $\phi(x,y)$, propagating in the dark dimension along with the graviton; its zero mode $\phi_0(x)$ is the field in \eqref{eq:fds}, and the radion is slaved to it along the valley of the potential, $r(x)\simeq0.04\,\phi_0(x)$ (through a potential $V\sim\Lambda\,(r(x)-0.04\,\phi_0(x))^2$). A massless bulk field comes with a KK tower,
\begin{equation}
\phi(x,y)=\phi_0(x)+\sum_{n\geq1}\phi_n(x)\,\psi_n(y)\,.
\end{equation}
{\it The evolving dark sector of} \cite{Bedroya:2025fwh} {\it therefore predicts, with nothing added, a second light tower in the dark sector!}

This tower is naturally another candidate for the dark matter, and two problems then solve each other. The KK gravitons produced from the SM brane at $T_i\sim1$~GeV must shed their mass before decaying back to the SM brane, and in two-derivative gravity (if we restrict to it) they cannot do so fast enough by splitting among themselves \cite{Lee:2026kkg,Langhoff:2026wiz}. Converting them into the $\phi$ tower, $G_n\to\phi_m\phi_l$, disposes of the gravitons and populates the $\phi$ tower. What makes the conversion fast is the following. In the five-dimensional effective theory the kinetic term of $\phi$ along $y$ is not tied to its four-dimensional kinetic term: the dark dimension is not a Lorentz-invariant five-dimensional vacuum due to profiles of bulk fields or fluxes along $y$, so the $\phi$ action takes the form
$$\tfrac12(\partial_\mu\phi)^2+\tfrac12 f(\phi,y)(\partial_y\phi)^2$$
with $f$ some function of $\phi$ and position. If the dark dimension is very homogeneous, as the scenario assumes, the $y$-dependence of $f$ is weak and $f(\phi,y)$ can be replaced by its $y$-average $\bar f(\phi)$, which on the zero mode is a propagation speed of $\phi$ along the dark dimension, $c_\phi^2=\bar f(\phi_0)$; subluminality requires $\bar f<1$. Taking $\bar f$ to be order one below one, the $\phi$ tower is lighter than the graviton tower level by level, $m_{\phi,n}\sim c_\phi\,n\,\mKK$, and $G_n\to\phi_m\phi_l$ with $m+l\sim n$ is open with an order-one release of energy. With the universal $1/\Mpl$ coupling of the graviton to the stress tensor of $\phi$, summed over the $\sim m/\mKK$ open channels (restoring $\Mpl$),
\begin{equation}
\Gamma(G_n\to\phi\phi)\,\sim\,(1-\bar f)^{5/2}\frac{m}{\mKK}\frac{m^3}{\Mpl^2}\,,
\end{equation}
where the prefactor comes from the $q^5$ term discussed in \cite{Lee:2026kkg,Langhoff:2026wiz}. Thus the GeV gravitons convert to the KK tower of $\phi$ on a time scale of seconds to days, much faster than their decay to SM fields or to lighter gravitons (if we restrict to two-derivative terms). The zero mode's four-dimensional kinetic term is untouched, so $\phi_0$ rolls as in Section~\ref{sec:fds}.

The daughters inherit the graviton production spectrum, so the tower is populated at GeV masses, and it then stays put: $\phi_n\to\phi_m G_{n-m}$ is closed because the $\phi$ tower is the lighter one; $\phi_n\to\phi_m\phi_l$ with KK momentum conserved sits essentially at threshold; and with a homogeneous dark dimension the KK-violating channels are negligible. The tower can thus be stable, unless we have extra ingredients to cause a cascade (see below). The cosmology is unaffected: the energy is handed from the gravitons to the $\phi$ quanta well before recombination, with only a tiny leak due to the converting gravitons' decay to SM fields. The dark matter is then a sterile spin-zero tower at the GeV scale, with no decay signal, whose one cosmological coupling is to the rolling $\phi_0$ through its masses, $m_{\phi,n}\sim c_\phi(\phi_0)\,n/\ell_5(\phi_0)$, so that
\begin{equation}
c'=-\frac{d\ln m_{\phi,n}}{d\phi_0}
  =\sqrt{\tfrac32}\,\sin\theta-\frac{d\ln c_\phi}{d\phi_0}\,.
\label{eq:cprime-two}
\end{equation}
The first term (with $\sin\theta\simeq0.04$) is the evolving dark dimension scenario of Section~\ref{sec:fds}. The second (setting $\theta=0$, with a stabilized dark dimension) is an alternative to it: $\bar f$ is set by the coefficients of the operators that couple $\partial\phi$ to the profiles along $y$, and these coefficients depend on the modulus $\phi$ like any other coupling of the internal effective theory. The dark-matter mass can therefore fade with the dark dimension rigidly stabilized and the radion heavy, in which case $\mKK$ remains fixed in time. Writing locally $\bar f(\phi)\simeq\bar f_0\,e^{-2c'\phi}$ gives $m_{\phi,n}\propto e^{-c'\phi_0}$, i.e.\ precisely \eqref{eq:fds} with $c'=-m'/m$. The FDS equations, the apparent phantom behavior and the dark fifth force $2c'^2$ are the same either way.

The $\phi_0$-dependence of $\bar f$ has a further consequence. Expanding $\bar f(\phi_0+\delta\phi)\,(\partial_y\delta\phi)^2$ to cubic order gives $c'\bar f\,\delta\phi\,(\partial_y\delta\phi)^2/\Mpl$, a cubic with derivatives only along $y$, which is not tied to the quadratic action and on the tower takes the form $\beta\,m^2/\Mpl$ of \cite{Gonzalo:2022jac} with $\beta\simeq c'\chi/2\sim10^{-2}$, $\chi$ the overlap of the KK profiles. Given a small inhomogeneity of the dark dimension, the tower then cascades, $\phi_n\to\phi_m\phi_l$, as the KK gravitons were assumed to in \cite{Gonzalo:2022jac,Obied:2023clp}: slowly, since $\beta$ is small, so that it sits at its production mass until about matter--radiation equality and fades afterwards along the attractor $m\propto t^{-2/7}$ by a factor $\sim30$, to tens of MeV today with kick velocities of a few km/s, conserving the dark-matter energy density and releasing nothing visible. The decaying-dark-matter phenomenology of \cite{Obied:2023clp}, the fading mass and the kicks, thus carries over to the $\phi$ tower unchanged, with the dark matter identified as the KK modes of the quintessence field rather than as the KK gravitons.
\end{appendices}

\end{document}